%% file: prl.tex
\documentclass[reprint,notitlepage,superscriptaddress,preprintnumbers,nofootinbib,amsmath,amssymb,aps,prl,nobibnotes]{revtex4-1}
\usepackage{graphicx}
\usepackage{dcolumn}
\usepackage{bm}
\usepackage{hyperref}

\input{macros.tex}

\begin{document}

\preprint{CERN-TH-2024-088}
\title{Light and Strange Vector Resonances from Lattice QCD at Physical Quark Masses}

\author{Peter Boyle}\affiliation{\bnl}\affiliation{\uoe}
\author{Felix Erben}\affiliation{\cern}\affiliation{\uoe}
\author{Vera G\"ulpers}\affiliation{\uoe}
\author{Maxwell T. Hansen}\affiliation{\uoe}
\author{Fabian Joswig}\affiliation{\uoe}
\author{\\Michael Marshall}\affiliation{\uoe}
\author{Nelson Pitanga Lachini}\email[e-mail: ]{np612@cam.ac.uk}\affiliation{\cambridge}\affiliation{\uoe}
\author{Antonin Portelli}\affiliation{\uoe}\affiliation{\cern}\affiliation{\rccs}

\begin{abstract}
We present the first \textit{ab initio} calculation at physical quark masses of scattering amplitudes describing the lightest pseudoscalar mesons interacting via the strong force in the vector channel. Using lattice quantum chromodynamics, we postdict the defining parameters for two short-lived resonances, the $\rho(770)$ and $K^*(892)$, which manifest as complex energy poles in $\pi \pi$ and $K \pi$ scattering amplitudes, respectively. The calculation proceeds by first computing the finite-volume energy spectrum of the two-hadron systems, and then determining the amplitudes from the energies using the L\"uscher formalism. The error budget includes a data-driven systematic error, obtained by scanning possible fit ranges and fit models to extract the spectrum from Euclidean correlators, as well as the scattering amplitudes from the latter. The final results, obtained by analytically continuing multiple parametrizations into the complex energy plane, are $M_\rho = 796(5)(50)\mev$, $\Gamma_\rho = 192(10)(31)\mev$, $M_{K^*} = 893(2)(54)\mev$ and $\Gamma_{K^*} = 51(2)(11)\mev$, where the subscript indicates the resonance and $M$ and $\Gamma$ stand for the mass and width, respectively, and where the first bracket indicates the statistical and the second bracket the systematic uncertainty.
\end{abstract}

\maketitle

\begin{figure*}[ht!]
    \centering
    \includegraphics[width=17.2cm]{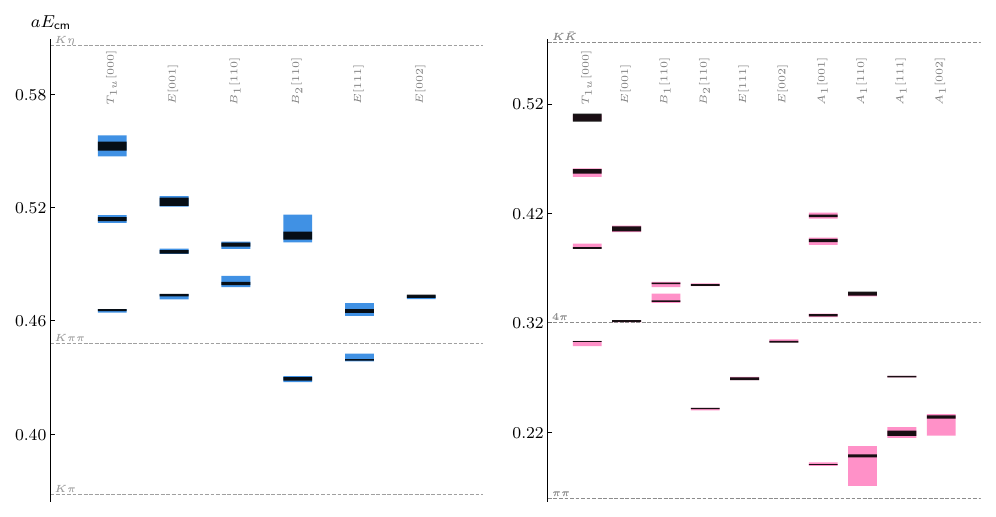}
    \caption{Finite-volume energy spectra extracted in this work for $K^*(892)$ (left panel) and $\rho(770)$ (right panel) quantum numbers and all irreducible representations considered. The span of the black rectangles represents statistical uncertainty and that of the colorful lighter ones is a data-driven systematic uncertainty determined from model averaging, as described in the main text. Relevant thresholds are indicated with gray lines.}
    \label{fig:spectrum}
\end{figure*}

\textit{Introduction} --- Over the past decade, precision has become increasingly crucial in particle-physics investigations of deviations between theory and experiment. In this vein, it is essential to reliably incorporate the strong force, defined by the theory of quantum chromodynamics ({QCD}), in all such predictions. In particular, resonances such as the $\rho (770)$ and $K^*(892)$, play an important role in the search for new physics beyond the standard model, because their detailed properties affect a wide range of observables, from the anomalous magnetic moment of the muon~\citep{\gminustwo} to heavy flavor weak decays that could reveal new $CP$-violating physics~\citep{\cpviolation}.

This Letter presents a state-of-the-art calculation of the properties of the aforementioned two resonances, each of which is clearly visible in experimental cross sections of their corresponding decay products: ${\pi \pi \to \rho(770) \to \pi \pi}$ (isospin ${I=1}$) and ${K \pi \to K^*(892) \to K \pi}$ (isospin ${I=1/2}$)~\citep{Protopopescu:1973sh,Estabrooks:1974vu,Estabrooks:1977xe,Aston:1987ir}. More precisely, such cross sections are used to extract partial-wave projected scattering amplitudes, where the resonances can be unambiguously characterized as poles in the complex energy plane. In this work, we extract the same partial-wave amplitudes and poles from a first-principles computation in the framework of lattice QCD. The only inputs are the QCD Lagrangian and the masses of pions and kaons as well as the omega baryon mass to set the fundamental energy scale of the theory.

In lattice QCD, the path integral defining the theory is evaluated numerically using Monte Carlo importance sampling. This is feasible only when the theory is defined in a discretized finite volume, with imaginary (Euclidean) time. As a result, all extracted correlation functions carry these modifications. As we detail in the following, it is possible to reliably extract finite-volume energies from such correlators which, while still depending on the lattice spacing and finite volume, do not carry any effects of the metric signature. In a second step, following the seminal work of L\"uscher and its many extensions~\citep{\AllFV}, these finite-volume energies can be related to the $\pi \pi \to \pi \pi$ and $K \pi \to K \pi$ partial-wave amplitudes.

A number of previous works have applied this work flow to investigate both the $\rho(770)$~\citep{\AllRho} and the $K^*(829)$~\citep{\AllKstar}. The calculation presented here is the first to directly use a physical implementation of the light and strange quark masses, in the definition of both explicit quarks within the scattering hadrons (valence quarks) and the quark-antiquark pairs arising as quantum fluctuations (sea quarks). It is additionally the first study of both resonances that uses the domain-wall quark discretization~\citep{Shamir:1993zy,Furman:1994ky,Brower:2012vk}, which is known to have desirable chiral symmetry properties, making it easier to reach physical masses without losing stability in the calculation.

The rest of this Letter is organized as follows: After briefly introducing our \textit{lattice setup}, we describe in detail how we have implemented the work flow outlined above. The first step is the \textit{spectrum determination}, in which we extract finite-volume energies for the channels of interest. Subsequently, we present our \textit{phase-shift determination} and explain our data-analysis procedure, designed to capture both the systematic and statistical uncertainties present in our data. By analytically continuing the phase shifts, we compute the resonance poles in the complex plane, which comprise the main results of this work. We close with some discussion and outlook concerning the future of scattering computations using numerical lattice QCD. This Letter is accompanied by a more detailed manuscript~\citep{our-long-manuscript}. The correlator data generated for this project are publicly available~\citep{boyle_2024_vy9x7-bzn92}.

\textit{Lattice setup} --- The computation was performed on a single RBC/UKQCD domain-wall-fermion (DWF) ensemble with geometry $(L/a)^3 \times (T/a)=48^3 \times 96$ and masses $m_\pi=\valueourpionmass$ and $m_K=\valueourkaonmass$, where $a$ is the lattice spacing, $T$ the temporal extent, and $L$ the spatial extent. The inverse lattice spacing on this ensemble has been previously determined to be $a^{-1}=\valueainverse$~\citep{RBC:2014ntl}. In all cases, values with physical units are determined by requiring the omega baryon mass to have its physical value, as is described in detail in Ref.~\citep{RBC:2014ntl}, where many other technical details of the ensemble are also given. (The exact values of $m_\pi$ and $m_K$ are new to this work, determined by combining our results for $a m_\pi$ and $a m_K$ with the previously determined lattice spacing.)

The key primary quantities determined on this ensemble are Euclidean two-point correlation functions of the form
\begin{equation}
    C_{ij}(t) \equiv \frac{1}{N_{t_s}} \sum_{t_s} \langle O_i(t+t_s) O_j(t_s)^{\dagger} \rangle \,,
    \label{eq:corr-matrix}
\end{equation}
where $O_i(t)$ is an operator, described in more detail below.
The correlation functions are averaged over $90$ gauge configurations,
$N_{t_s} = 96$ is the number of time slices, and the sum runs over all possible values, giving an additional average that takes advantage of the periodic boundary conditions in time to improve the statistical uncertainty.

\textit{Spectrum determination} --- Any operator with a given set of quantum numbers will generically have nonzero overlap with all finite-volume states sharing those quantum numbers. Relevant examples for this calculation are the vector bilinears:
\begin{align}
    \label{eq:bilinear_rho}
    O_{\rho}(\mathbf{P},t) &= a^3\sum_{\mathbf{x}} e^{-i \mathbf{P} \cdot \mathbf{x}} \bar{d}(x) {\bm \gamma} u(x) \,,   \\
    O_{K^{*}}\!(\mathbf{P},t) &= a^3\sum_{\mathbf{x}} e^{-i \mathbf{P} \cdot \mathbf{x}} \bar{s}(x) {\bm \gamma} u(x)  \,,
\label{eq:bilinear_Kstar}
\end{align}
where $x = (t, \textbf x)$ is a Euclidean four-vector and each operator is projected to definite spatial momentum $\mathbf{P}$, as shown. The total momentum satisfies ${\mathbf{P} = (2 \pi/L) \mathbf{d}}$, where $\mathbf{d} \in \mathbb{Z}^3$ is an integer three-vector.
Here, we have also introduced the Dirac spinor quark fields $u(x), d(x)$, and $s(x)$ as well as the spatial-component Dirac matrices $\bm \gamma = (  \gamma_x, \gamma_y, \gamma_z )$. In the infinite-volume context, the Dirac matrix ensures that the operator transforms as a component of a spatial three-vector. In the finite volume, this can be related to a definite row $r$, in an irreducible representation (\textit{irrep}) $\Lambda$, of the relevant finite-volume symmetry group. The latter depends on the value of $\textbf P$: For ${\textbf P = \textbf 0}$, the group is the 48-element octahedral group with parity, and for non-zero momenta a subgroup thereof. Details of the finite-volume group theory and corresponding operator construction are given in Refs.~\citep{\grouptheoryandops}.

In addition to the vector bilinears, we use nonlocal two-bilinear interpolators of the form
\begin{align}
    O_{K\pi}(x, y)    &= - K^+(x) \pi^0(y) + \sqrt{2} K^0(x) \pi^+(y) \,,  \\
    O_{\pi\pi}(x, y)  &=   \pi^+(x) \pi^0(y) -  \pi^0(x) \pi^+(y) \,,
    \label{pipikpiinterpolators}
\end{align}
where each local field on the right-hand side is a pseudoscalar quark bilinear with the quantum numbers of a kaon or pion, as indicated by the label, e.g.~${K^+(x) = \overline{s}(x)\gamma_5 u(x)}$.  -project these:
\begin{align}
    O_{MM'}(\mathbf p_1, \mathbf p_2,t) = a^6 \sum_{\mathbf x, \mathbf y} e^{-i(\mathbf p_1 \cdot \mathbf x+\mathbf p_2 \cdot \mathbf y)}O_{MM'}(x, y) \,,
\end{align}
where $MM' \in \{ K \pi, \pi \pi\}$, $x = (t, \textbf x)$, and $y = (t, \textbf y)$. The individual three-vectors $\textbf x$ and $\textbf y$ are projected to definite spatial momentum in the same way as in Eqs.~\eqref{eq:bilinear_rho} and \eqref{eq:bilinear_Kstar}. The resulting functions of two spatial momenta are then combined to form operators with definite total momentum $\textbf P$, as well as a definite irrep $\Lambda$ and row $r$.

This procedure leads to a set of interpolators with definite flavor, $\textbf P$, $\Lambda$, and $r$, a set of quantum numbers that we collectively denote by $\mathcal Q$. Such interpolators will generically overlap all states with the same $\mathcal Q$, and to obtain operators with improved overlap on a {\it specific} state, one requires a matrix of correlation functions, constructed from a set of operators as shown in Eq.~\eqref{eq:corr-matrix}. In the following, this is denoted by $C^{\mathcal Q}(t)$ to emphasize that it carries the same definite quantum numbers.

Such matrices can be efficiently evaluated using the method of distillation~\citep{Peardon:2009gh,Morningstar:2011ka}, which has been successfully applied in many lattice computations of resonance scattering processes~\citep{Lang:2011mn,Dudek:2012xn,Wilson:2015dqa,Bulava:2016mks,Andersen:2018mau,Erben:2019nmx,Prelovsek:2013ela,Wilson:2014cna,Wilson:2019wfr}. In this work we use exact distillation, taking advantage of the open-source implementation available in the Grid and Hadrons libraries~\citep{Boyle2015, Hadrons2023}.

To obtain the finite-volume energy spectrum, the final step is to solve a
generalized eigenvalue problem (GEVP) given by~\citep{Michael:1982gb,Michael:1985ne,Luscher:1990ck}
\begin{align}
    C^{\mathcal Q}(t) u^{\mathcal Q}_n(t)  = \lambda^{\mathcal Q}_n(t) C^{\mathcal Q}(t_0) u^{\mathcal Q}_n(t) \,,
    \label{eq:gevp}
\end{align}
where $n$ indexes the solution.
The GEVP eigenvalues are known to behave like
\begin{align}
    \lambda^{\mathcal Q}_n(t) =  Z_n^{\mathcal Q} \exp \! \left ( -E_n^{\mathcal Q} t \, \right ) \left [1 + \mathcal O(e^{- \Delta^{\mathcal Q}_n t}) \right ] \,,
    \label{eq:gevp-eigenvalue}
\end{align}
where $E_n^{\mathcal Q}$ denotes a finite-volume energy level in the spectrum and $\Delta^{\mathcal Q}_n > 0$ encodes the residual excited state contamination (see also the discussion in Ref.~\citep{Blossier:2009kd}). We obtain numerical estimates of $E_n^{\mathcal Q}$ by fitting a single exponential model to the $\lambda^{\mathcal Q}_n(t)$ data for sufficiently large $t$.

In this work, we consider five different values of spatial momentum in the finite-volume frame ($\mathbf d^2 = \textbf P^2 [L/(2 \pi)]^2 \in \{ 0,1,2,3,4\}$), which are useful as they effectively change the finite-volume geometry via a Lorentz contraction, leading to additional constraints on the scattering amplitudes. As shown in Fig.~\ref{fig:spectrum}, we are able to reliably extract a total of $13$ finite-volume energies across six irreps for the $K \pi$ scattering analysis. For the $\pi \pi$ analysis, we extract $21$ finite-volume energies across ten irreps.

\begin{figure*}[ht!]
    \centering
    \includegraphics[width=17.2cm]{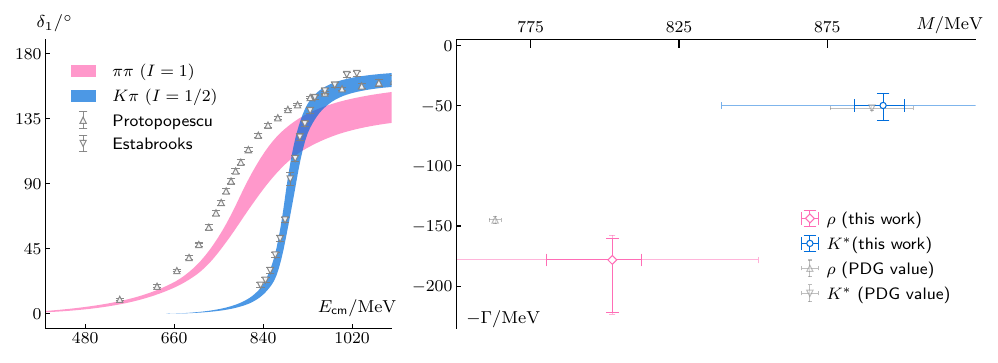}
    \caption{Left panel: results for the scattering phase shift for $K \pi \to K \pi$ and $\pi \pi \to \pi \pi$ (colorful), where the bands represent the uncertainties from statistical and data-driven systematic sources and \textit{do not} include other systematic errors, together with experimental phase-shift data (gray)~\citep{Protopopescu:1973sh,Estabrooks:1974vu,Estabrooks:1977xe,Aston:1987ir}. Right panel: resonance pole positions extracted from the second Riemann sheet, with statistical and data-driven systematics (colorful, larger error caps), and the other estimated systematics (colorful, fainter, smaller error caps) dominated by a conservative estimation of discretization effects due to the use of a single lattice spacing. In each case, all uncertainties were added in quadrature. The PDG averages (gray) come from unitarized chiral perturbation theory and dispersive analysis applied to experimental data~\citep{Pelaez:2020uiw, Colangelo:2001df, Garcia-Martin:2011nna,PDG2024}.}
    \label{fig:phaseshift}
\end{figure*}

\textit{Phase-shift determination} --- In the case that only a single flavor channel is relevant, the partial-wave projected scattering amplitude $t_{(\ell)}(p)$ can be expressed in terms of the scattering phase shift $\delta_\ell(p)$ via
\begin{equation}
    \tmat{\ell}(p) = \frac{1}{\cot \delta_\ell(p) - i} \,,
    \label{eq:scattering-phase}
\end{equation}
where $p$ is the magnitude of the center-of-mass frame momentum of one of the scatterers and $\ell$ denotes the orbital angular momentum. In the present case, we are interested in $\ell = 1$.

L\"uscher's formalism (and generalizations)~\citep{\AllFV} allows one to extract the scattering phase shift from the finite-volume energies. In the case where the $\ell = 3$ scattering phase shift is negligible, the general relation reduces to a simple algebraic expression
\begin{equation}
    \label{pwaveqcrootequation}
    \delta_1 ( p^{\mathcal Q}_n) = n \pi - \phi^{[\textbf P, \Lambda]}(p^{\mathcal Q}_n; L, m_1, m_2) \,,
\end{equation}
where $p^{\mathcal Q}_n$ is related to the extracted finite-volume energy via
\begin{equation}
    \sqrt{(E^{\mathcal Q}_n)^2 - \mathbf P^2} = \sqrt{m_1^2 + (p^{\mathcal Q}_n)^2 } + \sqrt{m_2^2 + (p^{\mathcal Q}_n)^2 } \,.
\end{equation}
Here $\phi^{[\textbf P, \Lambda]}$ is a geometric function (depending on the total momentum and the irrep: $\textbf P, \Lambda$), which can be readily computed to high precision, and $m_1,m_2 \in \{ m_\pi, m_K \}$ are the masses of the hadrons in the relevant scattering process.

Though the single-channel expression gives a direct determination of $\delta_1$ for each energy, we prefer to constrain the curve by fitting various models to the full dataset. One advantage is that this allows analytic continuation to the complex plane, to determine the resonance pole position. A second is that fit quality can be assessed across the entire work flow, \eg by examining how the quality of a given phase-shift fit depends on the time ranges used to determine the energies.

We denote a particular phase-shift model, with label $\sf m$, by the function $\delta_1^{\sf m}(p; \bm \alpha_{\sf m})$, where $\bm \alpha_{\sf m}$ is a vector of model parameters. Using this in Eq.~\eqref{pwaveqcrootequation} allows one to predict a set of model energies, denoted by $\{ E^{\mathcal Q, \sf m}_n(\bm \alpha_{\sf m}) \}$. These can be combined with numerically determined energies $\{ E^{\mathcal Q}_n \}$ to form a correlated chi-squared function $\chi^2_{\sf m}(\bm \alpha_{\sf m})$, which can be minimized with respect to $\bm \alpha_{\sf m}$ to determine the best-fit model parameters: $\bm \alpha_{\sf m}^\star$.

A key aspect of this work is a data-driven determination of the systematic uncertainty of $\bm \alpha_{\sf m}^\star$. This is achieved as follows:
First, we assign an Akaike information criterion ($\AIC$)~\citep{Akaike1974,Akaike1978,Borsanyi2021,Jay2021} to each energy-level fit
\begin{equation}
    [\AICc]_{n,k}^\mathcal{Q} =  [\chi^2]_{n,k}^\mathcal{Q} + 2 n^\mathrm{par} -  n_k^\mathrm{data},
    \label{eq:AICdef}
\end{equation}
where $n^\mathrm{par}=2$, the number of parameters entering the single-exponential fit, is fixed throughout. In addition to the state label $n$ and the quantum number label $\mathcal Q$, here we have included the label $k$ indexing all possible choices of the fit range $[\tmin, \tmax]$ used to extract $E^{\mathcal Q}_n$ from $\lambda_n^{\mathcal Q}(t)$. A given choice, denoted $[\tmin(k), \tmax(k)]$, leads to a value for $n_k^\mathrm{data} = \tmax(k) - \tmin(k) + 1$, and to a resulting value for the correlated chi-squared, $[\chi^2]_{n,k}^\mathcal{Q}$.

We then draw sets $\{ E^{\mathcal Q}_n\}_j, \ j=1, \ldots N$, each containing one representative fit (one particular $k$) for each energy level. For a given level $n$, the values of $\AICc$ determine the probability $[w_\mathrm{corr}]_{n,k}^{\mathcal{Q}} \propto e^{- [\AICc]_{n,k}^\mathcal{Q} /2}$ of a given fit result $k$ to be drawn into $\{ E^{\mathcal Q}_n \}_j$. Each set is then used to form a $\chi^2_{\sf m}(\bm \alpha_{\sf m})_j$ function as described above, yielding a particular $\bm \alpha_{{\sf m},j}^\star$. In this way, we obtain a distribution of model phase-shift parameters. In the final step, this distribution is weighted by an analogous exponentiation of the $\AIC_{\sf m}$ from the phase-shift fit. As the distribution was effectively weighted by $w_\mathrm{corr}$ in the fit-range sampling step described above, the final phase-shift parameters are, thus, distributed according to a total probability, given by a similar exponentiation of $\AIC_{\sf m} + \sum_n [\AICc]_{n,k}^\mathcal{Q}$. We assign a systematic error to the phase shift from the spread of this distribution. We have checked that each histogram contains fits with good $\chi^2_{\sf m} / n_\mathrm{d.o.f.}$ values, with the closest value to unity entering the distribution being $1.15$ for $\pi\pi$ scattering and $1.00$ for $K\pi$ scattering.

One can also use the spread of the $w_\mathrm{corr}$ distributions to give a separate systematic uncertainty for the energies themselves. The values of $\AIC_{\sf m}$ are not used here as the energies are intermediate results in the phase-shift calculation. The colorful rectangles in Fig.~\ref{fig:spectrum} indicate data-driven systematic uncertainties from this first $w_\mathrm{corr}$-weighted distribution.

\textit{Main results} --- We show the phase shifts for the $\rho$ and $K^*$ resonances in Fig.~\ref{fig:phaseshift}. Here, we sampled a total of $N=50\,000$ representative finite-volume energy fit samples and for each of them computed the phase-shift parameters of the Breit-Wigner and effective range models, which are the Breit-Wigner mass and coupling, and the scattering length and effective range, respectively. We also repeat such a procedure varying the minimum fit range and signal-to-noise ratio allowed in the fits to the GEVP data~\citep{our-long-manuscript}. A weighted histogram of these resonance parameters then lets us assign our final error estimates on them, which captures both systematic as well as statistical fluctuations in our data. Note that the statistical fluctuation of any single given fit is below our final quoted uncertainties. Our final results are then obtained by solving Eq.~\eqref{eq:scattering-phase} over all the analysis variations described above and lead to the pole-position parameters
\begin{align}
    \centering
    K^*(892)
    \begin{cases}
        M_{K^*}       &= 893(2)(8)(54)   \mev    \\ \nonumber
        \Gamma_{K^*}  &= 51(2)(11) (3)   \mev       \nonumber
    \end{cases}
\end{align}
and
\begin{align}
    \rho(770)
    \begin{cases}
        M_{\rho}       &= 796(5)(15)(48)    \mev \\  \nonumber
        \Gamma_{\rho}  &= 192(10)(28)(12)    \mev    \nonumber
    \end{cases}\,,
\end{align}
where $M$ and $\Gamma$ are the (pole) mass and width of the resonance, restpectively, indicated by the subscript. This is a symmetrized version of the result also depicted in Fig.~\ref{fig:phaseshift}. The first uncertainty comes from the statistical variation of the bootstrap samples, the second one comes from the data-driven procedure described above and the third one is an additional $6\%$ error associated with all other systematic uncertainties. The latter are dominated by the fact that we work only on a single lattice spacing but also include quark-mass mismatch, residual finite-volume effects, and the effects of inelastic thresholds such as $K\pi\pi$ and $\pi\pi\pi\pi$. It also includes the error stemming from the lattice scale setting, which is subdominant compared to the other sources of uncertainty. We note that the error in the left panel in Fig.~\ref{fig:phaseshift} does not include the additional $6\%$ uncertainty, and the experimental data points shown can be viewed as an indication of the potential discretization effects in our results. For a deeper discussion on the uncertainty budget, we refer the reader to~\citep{our-long-manuscript}.
Adding the systematic uncertainties in quadrature we arrive at
\begin{align}
    \centering
    K^*(892)
    \begin{cases}
        M_{K^*}       &= 893(2)(54)   \mev    \\ \nonumber
        \Gamma_{K^*}  &= 51(2)(11)    \mev       \nonumber
    \end{cases}
\end{align}
and
\begin{align}
    \rho(770)
    \begin{cases}
        M_{\rho}       &= 796(5)(50)    \mev \\  \nonumber
        \Gamma_{\rho}  &= 192(10)(31)    \mev    \nonumber
    \end{cases}\,.
\end{align}

\textit{Discussion and outlook} --- In this Letter, we have presented our calculation of the $\rho(770)$ and $K^*(892)$ resonance phase shift from lattice QCD at physical pion mass. This is the first physical-pion mass computation for the $K^*$ and the first one with physical pion mass and a dynamical strange quark for the $\rho$. Our result includes a full systematic error budget,
obtained from sets of underlying finite-volume energy levels, sampled by an $\AIC$ criterion. As mentioned before, the dominating uncertainty stems from our result being obtained from a single lattice spacing, which requires that we estimate the discretization effects directly as a percentage of the final results. This showcases that a crucial step forward would be to repeat this calculation on additional lattice spacings and take a continuum limit. In the right panel in Fig.~\ref{fig:phaseshift}, we compare our final results with the experimentally determined resonance pole positions, and find them to be in agreement within the quoted uncertainties at the $1\sigma$ level for $K^*$ and for the mass parameter of the $\rho$. The width of the $\rho$ agrees at $1.4\sigma$.

This computation is also a fœirst step toward any QCD process with $K \pi$ or $\pi \pi$ states present. One example is the muon $g-2$, where $\pi \pi$ scattering states help considerably in constraining the long-distance part and multiple groups have reported on progress with physical-point $\pi \pi$ scattering \cite{Bruno:2019nzm,Paul:2021pjz,Lahert:2021xxu,Lattice24:Miller,Lattice24:Lehner,RBC:2024fic,Lahert:2024vvu,Djukanovic:2024cmq}. Two other prominent such processes are $B \to K^* \ell^+ \ell^-$ and $B \to \rho \ell \nu$. Given recent experimental results~\citep{LHCb:2022qnv}, a lattice result on these decays will be an important input for improved tests of the standard model. Existing lattice calculations on these decays~\citep{Bowler:2004zb,Flynn:2008zr,Horgan:2013hoa} have used the narrow-width approximation in which the $\rho$ or $K^*$ are assumed to be a QCD-stable state. Some progress in going beyond this approximation has recently been reported in Ref.~\citep{Leskovec:2022ubd}, further demonstrating that lattice QCD is reaching the era where such computations, with resonant final states, are realistic.

\bigskip

The data entering this publication was generated using the openly available lattice QCD software packages Grid~\citep{Boyle2015} and Hadrons~\citep{Hadrons2023}. The supporting data for this article are openly available from the CERN document server~\citep{boyle_2024_vy9x7-bzn92}.

\bigskip

\textit{Acknowledgments}---The authors thank the members of the RBC and UKQCD Collaborations for the helpful discussions and suggestions. N. P. L., F. E. and A. P. kindly thank Mike Peardon for the invaluable discussions. N. P. L. additionally thanks André Baião Raposo for the discussions. This work used the DiRAC Extreme Scaling service (Tursa / Tesseract) at the University of Edinburgh, managed by the Edinburgh Parallel Computing Centre on behalf of the STFC DiRAC HPC Facility. The DiRAC service at Edinburgh was funded by BEIS, UKRI and STFC capital funding, and STFC operations grants. DiRAC is part of the UKRI Digital Research Infrastructure. P. B. has been supported in part by the U.S. Department of Energy, Office of Science, Office of Nuclear Physics under the Contract No. DE-SC-0012704 (BNL). P. B. has also received support from the Royal Society Wolfson Research Merit award WM/60035. M. M. gratefully acknowledges support from STFC in the form of a fully funded Ph.D. studentship. A. P. and F. E. received funding from the European Research Council (ERC) under the European Union's Horizon 2020 research and innovation program under Grant Agreement No. 757646. N. P. L. and A. P. received funding from the European Research Council (ERC) under the European Union's Horizon 2020 research and innovation programme under grant agreement No. 813942. N. P. L. also acknowledges support from the United Kingdom Science and Technology Facilities Council (STFC) [Grants No. ST/T000694/1, ST/X000664/1]. F. E. has received funding from the European Union's Horizon Europe research and innovation programme under the Marie Sk\l{}odowska-Curie Grant Agreement No. 101106913. M. T. H. and F. J. are supported by UKRI Future Leader Fellowship No. MR/T019956/1. A. P., M. T. H., V. G. and F. E. were supported in part by United Kingdom STFC Grant No. ST/P000630/1, and No. A. P., M. T. H. and V. G. additionally by UK STFC grants No. ST/T000600/1 and No. ST/X000494/1. A. P. is additionally partially supported by a long-term Invitational Fellowship from the Japan Society for the Promotion of Science.

\bibliographystyle{apsrev4-1}
\bibliography{scattering}

\newpage

\end{document}

%% file: macros.tex
\newcommand{\uoe}{School of Physics and Astronomy,
University of Edinburgh, Edinburgh EH9 3JZ, United Kingdom}
\newcommand{\bnl}{Physics Department, Brookhaven National Laboratory, Upton, New York 11973, USA}
\newcommand{\cern}{CERN, Theoretical Physics Department, Geneva, Switzerland}
\newcommand{\cambridge}{DAMTP, University of Cambridge, Centre for Mathematical Sciences, Wilberforce Road, Cambridge, CB3 0WA, United Kingdown}
\newcommand{\rccs}{RIKEN Center for Computational Science, Kobe 650-0047, Japan}

\newcommand{\gminustwo}{Aoyama:2020ynm}
\newcommand{\cpviolation}{Buras:1998raa}

\newcommand{\grouptheoryandops}{Elliot1979a, Moore:2005dw, Thomas2012, Foley2011, Gockeler2012a, Prelovsek2017, Detmold2024b}
\newcommand{\AllRho}{CP-PACS:2007wro,CS:2011vqf,Feng:2010es,Lang:2011mn,Pelissier:2012pi,Dudek:2012xn,Wilson:2015dqa,Bali:2015gji,Bulava:2016mks,Fu:2016itp,Andersen:2018mau,Erben:2019nmx,Alexandrou:2017mpi,ExtendedTwistedMass:2019omo,Fischer:2020yvw}
\newcommand{\AllKstar}{Fu:2012tj,Prelovsek:2013ela,Wilson:2014cna,Wilson:2019wfr,Bali:2015gji,Brett:2018jqw,Rendon:2020rtw}
\newcommand{\AllFV}{Luscher:1985dn,Luscher:1986pf,Luscher:1990ux,Rummukainen:1995vs,Kim:2005gf,Fu:2011xz,Leskovec:2012gb,Bernard:2010fp,Doring:2011vk,Briceno:2012yi,Hansen:2012tf,Briceno:2014oea}

\DeclareMathOperator{\valueainverse}{ 1.7295(38)~\mathrm{GeV} } 
\DeclareMathOperator{\valueourpionmass}{ 138.5(2)~\mathrm{MeV} } 
\DeclareMathOperator{\valueourkaonmass}{ 498.9(4)~\mathrm{MeV} }

\usepackage[dvipsnames]{xcolor}
\usepackage{color}
\usepackage{amsmath,amssymb,braket,slashed,mathrsfs,dsfont}
\usepackage{pifont}
\usepackage{braket}
\usepackage[compat=1.1.0]{tikz-feynhand}
\usepackage{mathtools}
\usepackage{subcaption}
\usepackage[mathscr]{euscript}
\usepackage{multirow}
\usepackage[normalem]{ulem} 

\hypersetup{colorlinks, linkcolor = [rgb]{0,0.0,0.75}, citecolor = [rgb]{0,0.0,0.75}, urlcolor = [rgb]{0,0.0,0.75}}

\newcommand{\eg}{\textit{e.g.}~}

\DeclareMathOperator{\mev}{\!~MeV}

\DeclareMathOperator{\AIC}{\mathrm{AIC}}

\newcommand{\tmin}{t_\mathrm{min}}
\newcommand{\tmax}{t_\mathrm{max}}
\newcommand{\AICc}{\mathrm{AIC}_\mathrm{corr}}

\newcommand{\tmat}[1]{t_{#1}} 

\makeatletter
\newcommand{\thickhline}{%
    \noalign {\ifnum 0=`}\fi \hrule height 1pt
    \futurelet \reserved@a \@xhline
}
\newcolumntype{"}{@{\hskip\tabcolsep\vrule width 1pt\hskip\tabcolsep}}
\makeatother